\documentclass[twocolumn,showpacs,showkeys,amsmath,amssymb]{revtex4}

\usepackage{graphicx}
\usepackage{bm} 
\usepackage{subfigure}
\usepackage[T1]{fontenc}

\begin{document}
 
\title{Characteristic form of boost-invariant and cylindrically non-symmetric hydrodynamic equations
\footnote{\\ Research supported by the Polish State Committee for
Scientific Research, grant 2P03B~05925}}

\author{Miko{\l }aj Chojnacki} 
\email{Mikolaj.Chojnacki@ifj.edu.pl}
\affiliation{The H. Niewodnicza\'nski Institute of Nuclear Physics, 
Polish Academy of Sciences, PL-31342 Krak\'ow, Poland}

\author{Wojciech Florkowski} 
\email{Wojciech.Florkowski@ifj.edu.pl}
\affiliation{Institute of Physics, \'Swi\c{e}tokrzyska Academy,
ul.~\'Swi\c{e}tokrzyska 15, PL-25406~Kielce, Poland} 
\affiliation{The H. Niewodnicza\'nski Institute of Nuclear Physics, 
Polish Academy of Sciences, PL-31342 Krak\'ow, Poland}

\date{March 27, 2006}

\begin{abstract}
It is shown that the boost-invariant and cylindrically non-symmetric hydrodynamic equations for baryon-free matter may be reduced to only two coupled differential equations. In the case where the system exhibits the cross-over phase transition, the standard numerical methods may be applied to solve these equations and the proposed scheme allows for a very convenient analysis of the cylindrically non-symmetric hydrodynamic expansion.
\end{abstract}



\maketitle 

\section{Introduction}
\label{sect:Intro}

The success of the hadron production models based on the relativistic hydrodynamics \cite{Teaney:2001av,Huovinen:2001cy,Huovinen:2002fp,Kolb:2002ve,Hirano:2002ds,Hirano:2004rs,Hirano:2005wx} in describing the RHIC data brings the evidence that the hot and dense matter produced at RHIC behaves like an almost perfect fluid \cite{Heinz:2005zg}. This situation, in turn, triggers interest in the further development of the hydrodynamic models, especially in the more detailed studies of the dissipative effects \cite{Chaudhuri:2005ea,Heinz:2005bw,Baier:2006um}. In this paper we consider the hydrodynamics of the perfect fluid and show that the boost-invariant and cylindrically non-symmetric relativistic hydrodynamic equations for baryon-free matter may be comfortably reduced to only two coupled differential equations. The effects of the cross-over phase transition may be included in this scheme by the use of the temperature dependent sound velocity $c_s(T)$. As long as the function $c_s(T)$ satisfies the stability condition against the shock formation, the resulting equations may be solved with the help of standard numerical methods and the proposed scheme allows for a very convenient analysis of the cylindrically non-symmetric hydrodynamic expansion. The proposed method may be used to describe the evolution of matter produced in the central region of ultra-relativistic heavy-ion collisions, such as investigated in the present RHIC or future LHC experiments. 

The presented formalism is a direct generalization of the approach introduced by Baym, Friman, Blaizot, Soyeur, and Czyz \cite{Baym:1983sr}. In Ref. \cite{Baym:1983sr} the boost-invariant and cylindrically symmetric systems were considered, and the numerical calculations were performed only for the case of the constant sound velocity, $c_s^2 = \frac{1}{3}$. The effects of the temperature dependent sound velocity were included in Ref. \cite{Chojnacki:2004ec}, where also the possibility of the non-zero initial transverse flow was considered in setting up the initial conditions for the hydrodynamic equations. In this paper we further relax the assumptions of Ref. \cite{Baym:1983sr} by considering the cylindrically non-symmetric systems. 

\section{Hydrodynamic equations}
\label{sect:HE}

\begin{figure}[t]
\begin{center}
\includegraphics[angle=0,width=0.3\textwidth]{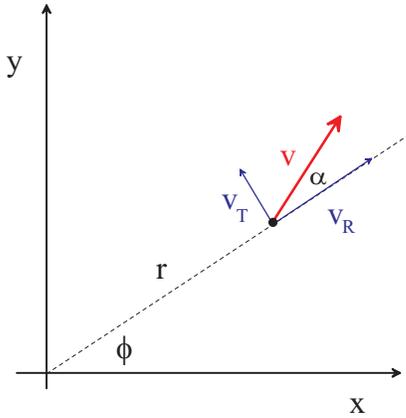}
\end{center}
\caption{Decomposition of the flow velocity vector in the plane $z=0$. In our approach
we use the magnitude of the flow $v$ and the angle $\alpha$ as two independent quantities.}
\label{fig:alpha}
\end{figure}

The relativistic hydrodynamic equations of the perfect fluid follow from the
energy-momentum conservation law and the assumption of the local equilibrium. For baryon-free matter they read
\begin{eqnarray}
u^\mu \partial_\mu (T u^\nu) &=& \partial^\nu T, 
\label{h1} \\
\partial_\mu (s u^\mu) &=& 0, 
\label{h2}
\end{eqnarray}
where $T$ is the temperature, $s$ is the entropy density, and $u^\mu = \gamma(1,{\bf v})$ is the four-velocity of the fluid. Due to the normalization of the four-velocity, only three out of four equations appearing in (\ref{h1}) are independent. The assumption of the boost-invariance introduces another constraint, hence, in this case Eqs. (\ref{h1}) and (\ref{h2}) are reduced to three independent equations \cite{Dyrek:1984xz}
\begin{eqnarray}
\frac{\partial }{\partial t}\left( rts\gamma \right) +\frac{\partial }{\partial r}
\left( rts\gamma v\cos \alpha \right) + \frac{\partial }{\partial
\phi }\left( ts\gamma v\sin \alpha \right)  &=&0, 
\nonumber \\
\frac{\partial }{\partial t}\left( rT\gamma v\right) +
r\cos \alpha \frac{\partial }{\partial r}\left( T\gamma \right) 
+\sin \alpha \frac{\partial }{\partial \phi }\left( T\gamma \right)  &=&0, 
\nonumber \\
T\gamma ^{2}v\left( \frac{d\alpha }{dt}+\frac{v\sin \alpha }{r}\right) -\sin
\alpha \frac{\partial T}{\partial r}
+\frac{\cos \alpha }{r}\frac{\partial T}{\partial \phi } &=&0. \nonumber \\
\label{wfdyr}
\end{eqnarray}
Here $t, r\!\!=\!\!\sqrt{x^2+y^2}$, and $\phi=\hbox{tan}^{-1} (y/x)$ are the time and space coordinates which parameterize the plane $z=0$ (due to the boost-invariance of the system, the values of all physical quantities at $z \neq 0$ may be obtained directly by the Lorentz boosts). The quantity $v$ is the magnitude of the fluid velocity, $\gamma =\left(1-v^2\right)^{-\frac{1}{2}}$ is the corresponding Lorentz factor, while $\alpha$ is the function describing direction of the flow, $\alpha=\hbox{tan}^{-1}(v_T/v_R)$, see Fig.~\ref{fig:alpha}. The differential operator $\frac{d}{dt}$ in (\ref{wfdyr}) is defined by the formula
\begin{equation}
\frac{d}{dt} = \frac{\partial}{\partial t} 
+  v \cos\alpha \frac{\partial}{\partial r} 
+ \frac{v \sin\alpha}{r} \frac{\partial}{\partial \phi}.
\end{equation}

Equations (\ref{wfdyr}) are three equations for four unknown functions: $T$, $s$, $v$, and $\alpha$. To form a closed system of equations they should be supplemented by the equation of state, i.e., by the relation connecting $T$ and $s$. Alternatively, the equation of state may be included by the use of the temperature dependent sound velocity 
\begin{equation}
c_s^2(T) = \frac{\partial P}{\partial \epsilon} = 
\frac{s}{T}\frac{\partial T}{\partial s},
\label{cs}
\end{equation}
and by the use of the potential $\Phi(T)$ defined by the equation
\begin{equation}
d\Phi=\frac{d\ln T}{c_s}=c_s d\ln s.
\label{phi}
\end{equation}

\begin{figure}[t]
\begin{center}
\includegraphics[angle=0,width=0.45\textwidth]{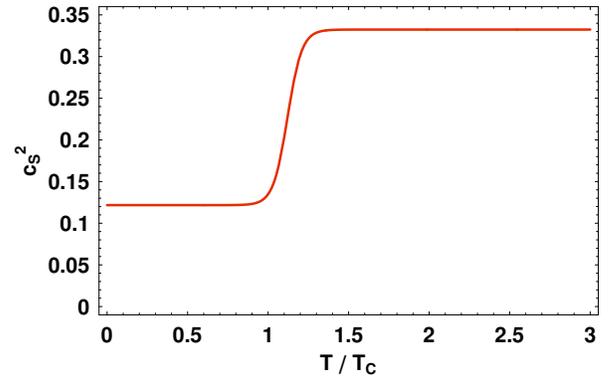}
\end{center}
\caption{Sound velocity squared shown as a function of the temperature (with the critical temperature $T_c =$ 170 MeV). The presented form is obtained by matching the lattice results with the resonance gas model calculation \cite{Mohanty:2003va,Chojnacki:2004ec}. The cross-over phase transition is considered, hence no sudden decrease (dip) of the sound velocity is observed at $T=T_c$.  }
\label{fig:cs2}
\end{figure}

The form of the function $c_s^2(T)$ used in our calculations is shown in Fig. \ref{fig:cs2}. This form, with $T_c$ = 170 MeV, follows from matching the lattice results with the hadron resonance gas calculation \cite{Mohanty:2003va,Chojnacki:2004ec}. We note that integration of Eq. (\ref{phi}) allows us to express $\Phi$ in terms of the temperature, $\Phi = \Phi_T(T)$, or to express temperature in terms of $\Phi$, $T=T_\Phi(\Phi)$. The functions $\Phi_T(T)$ and $T_\Phi(\Phi)$, corresponding to the sound velocity function shown in Fig. \ref{fig:cs2}, are presented and discussed in more detail in Ref. \cite{Chojnacki:2004ec}. The condition against the shock formation has the form \cite{Baym:1983sr,Blaizot:1987cc}
\begin{equation}
 1 - c_s^2 + c_s T \frac{dc_s}{dT} 
= {d \over dT} \left({s \, c_s \over T } \right) \ge 0.
\label{stab1}
\end{equation}
We note that this condition is naturally fulfilled in our case, since the derivative of $c_s(T)$ is always positive.

With the help of the substitutions:
\begin{equation}
v = \tanh \theta, \quad a_\pm = \exp(\Phi \pm \theta),
\label{subst}
\end{equation}
the sum and the difference of the first two equations in (\ref{wfdyr}) may be rewritten as
\begin{eqnarray}
&& \frac{\partial a_{\pm }}{\partial t}+\frac{\left( v\pm c_{s}\right) }{%
\left( 1\pm c_{s}v\right) }\cos \alpha \frac{\partial a_{\pm }}{\partial r}+%
\frac{\left( v\pm c_{s}\right) }{\left( 1\pm c_{s}v\right) }\frac{\sin
\alpha }{r}\frac{\partial a_{\pm }}{\partial \phi } 
\nonumber \\
&& -\,\frac{c_{s}v}{\left( 1\pm c_{s}v\right) }\left( \sin \alpha \frac{%
\partial \alpha }{\partial r}-\frac{\cos \alpha }{r}\frac{\partial \alpha }{%
\partial \phi }\right) \,a_{\pm }
\nonumber \\
&& 
+ \,\frac{c_{s}}{\left( 1\pm c_{s}v\right) }%
\left[ \frac{1}{t}+\frac{v\cos \alpha }{r}\right] \,a_{\pm } =0,
\label{apmeq}
\end{eqnarray}
while the third equation in (\ref{wfdyr}) is
\begin{eqnarray}
\frac{\partial \alpha }{\partial t} &=&
\frac{\left( 1-v^{2}\right) c_{s}}{v}
\left( \sin \alpha \frac{\partial \Phi }{\partial r}-\frac{\cos \alpha }{r}%
\frac{\partial \Phi }{\partial \phi }\right) 
\nonumber \\
& & 
- v \left( \cos \alpha \frac{\partial \alpha }{\partial r}
+\frac{\sin \alpha }{r}\frac{\partial \alpha }{\partial
\phi }+\frac{\sin \alpha }{r}\right).
\label{alphaeq}
\end{eqnarray}
Equations (\ref{apmeq}) and (\ref{alphaeq}) are three equations for three unknown functions: $a_+(t,r,\phi)$, $a_-(t,r,\phi)$, and $\alpha(t,r,\phi)$. We note, that the velocity $v$ and the potential $\Phi$ are functions of $a_+$ and $a_-$, 
\begin{equation}
v = {a_+ - a_- \over a_+ + a_-}, \quad \Phi = \frac{1}{2} \ln (a_+ a_-).
\label{vPhi}
\end{equation}
Similarly, the sound velocity appearing in Eqs. (\ref{apmeq}) and (\ref{alphaeq}) may be also represented as a function of $a_+$ and $a_-$, 
\begin{equation}
c_s(T) = c_s\left\{T_\Phi\left[\frac{1}{2} \ln (a_+ a_-)\right] \right\}.
\label{csapam}
\end{equation}

\section{Boundary conditions}
\label{sect:bcond}

\subsection{With cylindrical symmetry}
\label{sect:symbcond}

In the special case of the cylindrical symmetry all terms in Eq. (\ref{alphaeq}) vanish 
($\partial \Phi/\partial \phi =0$ and $\alpha =0$, since we do not consider the possibility that the system rotates as a whole with $\alpha=\pi/2$), whereas Eqs. (\ref{apmeq}) are reduced  to Eqs. (2.24) of Ref. \cite{Baym:1983sr}
\begin{equation}
\frac{\partial a_{\pm }}{\partial t}+\frac{\left( v\pm c_{s}\right) }{%
\left( 1\pm c_{s}v\right) }\frac{\partial a_{\pm }}{\partial r}+\,\frac{c_{s}%
}{\left( 1\pm c_{s}v\right) }\left[ \frac{1}{t}+\frac{v}{r}\right] 
\,a_{\pm} =0.
\label{apmbaym}
\end{equation}
Due to the symmetry reasons, the velocity $v$ should vanish at $r=0$,
\begin{equation}
v(t,r=0) =0.
\label{boundv}
\end{equation}
This condition is achieved by demanding that the two functions $a_+(t,r)$ and $a_-(t,r)$ may be expressed in terms of a single function $a(t,r)$ with the help of the prescription
\begin{eqnarray}
a_+(t,r) &=& a(t,r),\quad r>0, \nonumber \\
a_-(t,r) &=& a(t,-r),\quad r<0.
\label{adefbaym}
\end{eqnarray}
The ansatz (\ref{adefbaym}) and the structure of Eqs. (\ref{apmbaym}) indicate that the equation for $a_-(t,r)$ may be obtained from the equation for $a_+(t,r)$ by the substitution: $r \to -r$. This observation further means that two equations (\ref{apmbaym}) may be reduced to a single equation for the function $a(t,r)$ with the range of the variable $r$ extended to negative values,
\begin{equation}
0=\frac{\partial a}{\partial t}
+\frac{\left( v + c_{s}\right) }{\left( 1 + c_{s}v\right) }
\frac{\partial a}{\partial r}+\,\frac{c_{s}}{\left( 1 + c_{s}v\right) }
\left[ \frac{1}{t}+\frac{v}{r}\right] \,a.
\label{abaym}
\end{equation}
Furthermore, the ansatz (\ref{adefbaym}) and Eq. (\ref{vPhi}) automatically yield the desired boundary condition for the temperature
\begin{equation}
\left. \frac{\partial T(t,r)}{\partial r} \right|_{r=0} = 0.
\label{boundT}
\end{equation}
The scheme presented above is the basis of the approach used in Ref. \cite{Baym:1983sr}. Below we develop the analogous scheme for cylindrically non-symmetric systems. 

\begin{figure}[t]
\begin{center}
\includegraphics[angle=0,width=0.45\textwidth]{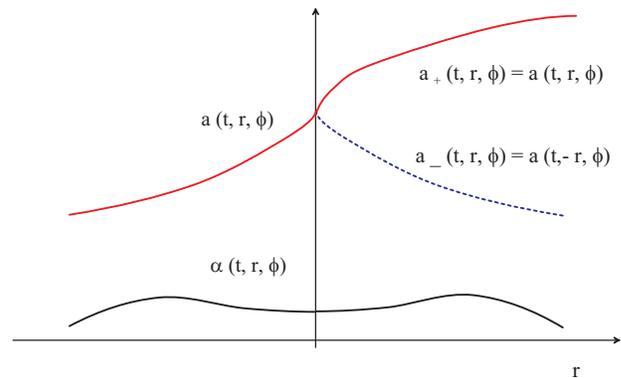}
\end{center}
\caption{Construction of the functions $a_+ (t,r,\phi)$ and $a_-(t,r,\phi)$ with the help of the function $a(t,r,\phi)$. The function $\alpha(t,r,\phi)$ is symmetrically extended to negative values of $r$. }
\label{fig:rsym}
\end{figure}

\subsection{Without cylindrical symmetry}
\label{sect:symbcond}

In this paper we consider the collisions of identical nuclei with atomic number $A$  which collide moving initially along the $z$-axis. The impact parameter ${\bf b}$ points in the $x$-direction. The center of the first nucleus in the transverse plane is placed at ${\bf x}_1 = (x_1,y_1)=(-b/2,0)$, and of the second nucleus at ${\bf x}_2 = (x_2,y_2)=(b/2,0)$. In such a case, similarly to the symmetric case considered above, we also require that the magnitude of the flow vanishes at ${\bf x} = (x,y)=(0,0)$, i.e., for $r=0$ we have
\begin{equation}
v(t,r=0,\phi) =0.
\label{boundvphi}
\end{equation}
This condition is fulfilled naturally by the ansatz analogous to Eq. (\ref{adefbaym}), namely
\begin{eqnarray}
a_+(t,r,\phi) &=& a(t,r,\phi),\quad r>0, \nonumber \\
a_-(t,r,\phi) &=& a(t,-r,\phi),\quad r<0.
\label{adefwe}
\end{eqnarray}
We supplement the ansatz (\ref{adefwe}) by the definition of the function $\alpha(t,r,\phi)$ for the negative arguments, see Fig. \ref{fig:rsym},
\begin{eqnarray}
\alpha(t,-r,\phi) &=& \alpha(t,r,\phi), \quad r > 0.
\label{alphadefwe}
\end{eqnarray}
With the help of the definitions (\ref{adefwe}) and (\ref{alphadefwe}), Eqs. (\ref{apmeq}) may be reduced to a single equation for the function $a(t,r,\phi)$,  
\begin{eqnarray}
&& \frac{\partial a}{\partial t}+\frac{\left( v + c_{s}\right) }{%
\left( 1 + c_{s}v\right) }\cos \alpha \frac{\partial a}{\partial r}+%
\frac{\left( v + c_{s}\right) }{\left( 1 + c_{s}v\right) }\frac{\sin
\alpha }{r}\frac{\partial a}{\partial \phi } 
\nonumber \\
&& -\,\frac{c_{s}v}{\left( 1 + c_{s}v\right) }\left( \sin \alpha \frac{%
\partial \alpha }{\partial r}-\frac{\cos \alpha }{r}\frac{\partial \alpha }{%
\partial \phi }\right) \,a
\nonumber \\
&& + \,\frac{c_{s}}{\left( 1 + c_{s}v\right) }%
\left[ \frac{1}{t}+\frac{v\cos \alpha }{r}\right] \,a =0,
\label{aeq}
\end{eqnarray}
Here, similarly to the cylindrically symmetric case, the range of the variable $r$ is extended to negative values. Eq. (\ref{aeq}) should be solved together with Eq. (\ref{alphaeq}), where the range of $r$ may be also extended trivially to negative values (definitions (\ref{adefwe}) and (\ref{alphadefwe}) imply that this equation is in fact invariant under transformation: $r \to -r$). The use of the polar coordinates in the transverse plane requires also that all functions at $\phi=0$ and $\phi=2\pi$ are equal: $a(t,r,0) = a(t,r,2 \pi), \alpha(t,r,0) = \alpha(t,r,2 \pi)$.
We also note that Eqs. (\ref{adefwe}) and (\ref{alphadefwe}) yield the following boundary conditions for the temperature and the function $\alpha(t,r,\phi)$, 
\begin{equation}
\left. \frac{\partial T(t,r,\phi) }{\partial r} \right|_{r=0} = 0, \quad
\left. \frac{\partial \alpha(t,r,\phi)}{\partial r} \right|_{r=0} = 0.
\label{boundTalpha}
\end{equation}
In addition, multiplying the equation for $a_+$ in (\ref{apmeq}) by $a_-$ and subtracting the equation for $a_-$ multiplied by $a_+$ one finds
\begin{equation}
\left. \frac{1}{r} \frac{\partial T(t,r,\phi) }{\partial \phi} \right|_{r=0} = 0.
\label{boundTphi}
\end{equation}

\section{Initial conditions}

\subsection{Temperature}

In the following we assume that the hydrodynamic evolution starts at a typical time $t=t_{0}=1$ fm.  We assume also that the initial temperature profile is connected with the number of participating nucleons
\begin{equation}
T(t_0,x,y) = \hbox{const} \left( \frac{dN_p}{dx\,dy} \right)^{1/3},
\label{Tt0}
\end{equation}
where we use the formula \cite{Teaney:2001av}
\begin{eqnarray}
& &\!\!\!\!\!\!\!\! \frac{dN_p}{dx\,dy} = 
T_A\left(\frac{\bf b}{2}+ {\bf x} \right)
\left\{1\! -\! \left[1\! -\! \frac{\sigma_{\rm in}}{A} \, T_A\left(-\frac{\bf b}{2}+ {\bf x} \right)
\right]^A \right\} \nonumber \\
& &\!\!\!\!\!\!\!\! + \, T_A\left(-\frac{\bf b}{2}+ {\bf x} \right)
\left\{1\! -\! \left[1\! -\! \frac{\sigma_{\rm in}}{A} \, T_A\left(\frac{\bf b}{2}+ {\bf x} \right)
\right]^A \right\}. \nonumber \\
\label{dNp}
\end{eqnarray}
In Eq. (\ref{dNp}) $\sigma_{\rm in}$ = 40 mb is the inelastic nucleon-nucleon cross section and $T_A\left(x,y\right)$ is the nucleus thickness function
\begin{equation}
T_A(x,y) = \int dz \, \rho\left(x,y,z\right).
\label{TA}
\end{equation}
Here $\rho(r)$ is the nuclear density profile given by the Woods-Saxon function with a conventional choice of the parameters used for the gold nucleus: 
\begin{eqnarray}
\rho_0 &=& 0.17 \,\hbox{fm} ^{-3},  \nonumber \\
r_0 &=& (1.12 A^{1/3} -0.86 A^{-1/3}) \,\hbox{fm},  \nonumber \\
a &=& 0.54 \,\hbox{fm}, \quad A = 197.
\label{woodssaxon}
\end{eqnarray}
The idea to use Eq. (\ref{Tt0}) follows from the assumption that the initially produced entropy density  $\sigma(t_0,x,y)$ is proportional to the number of the nucleons participating in the collision at the position $(x,y)$. Since the considered systems are initially very hot (with the temperature exceeding the critical temperature $T_c$), they may be considered as systems of massless particles, where the entropy density is proportional to the third power of the temperature. In this way we arrive at Eq. (\ref{Tt0}).

\subsection{Velocity}

The initial form of the functions $v(t,r,\phi)$ and $\alpha(t,r,\phi)$ is
\begin{eqnarray}
v(t_0,r,\phi) &=& v_{\,0}(r) =  \frac{ H_0 r}{\sqrt{1 + H_0^2 r^2}}, \nonumber \\
\alpha(t_0,r,\phi) &=& 0,
\label{initv}
\end{eqnarray}
where $H_0$ is a parameter defining the initial transverse flow formed in the pre-equilibrium stage. In the present calculations we use a very small value $H_0$ =0.001 fm$^{-1}$. We note that the finite value of the initial transverse flow makes all the terms on the right-hand-side of Eq. (\ref{alphaeq}) finite for $r > 0$. At $r=0$ these terms are also finite due to the boundary conditions (\ref{boundvphi}) and (\ref{boundTphi}). 

The two initial conditions, Eqs. (\ref{Tt0}) and (\ref{initv}), may be included in the
initial form of the function $a(r)$ if we define
\begin{equation}
a(t=t_0,r,\phi) =
a_T(r,\phi) \frac{\sqrt{1 + v^{\,0}(r)}}{\sqrt{1 - v^{\,0}(r)} },
\label{initaT}
\end{equation}
where
\begin{equation}
a_T(r,\phi) = \hbox{exp}
\left\{\Phi_T
\left[
\hbox{const} \,\, \left( \frac{dN_p}{dx\,dy} \right)^{1/3} \right]
\right\}.
\label{aT}
\end{equation}

\section{Results}

\begin{figure*}[t!]
\begin{center}
\subfigure{\includegraphics[angle=0,width=0.95\textwidth]{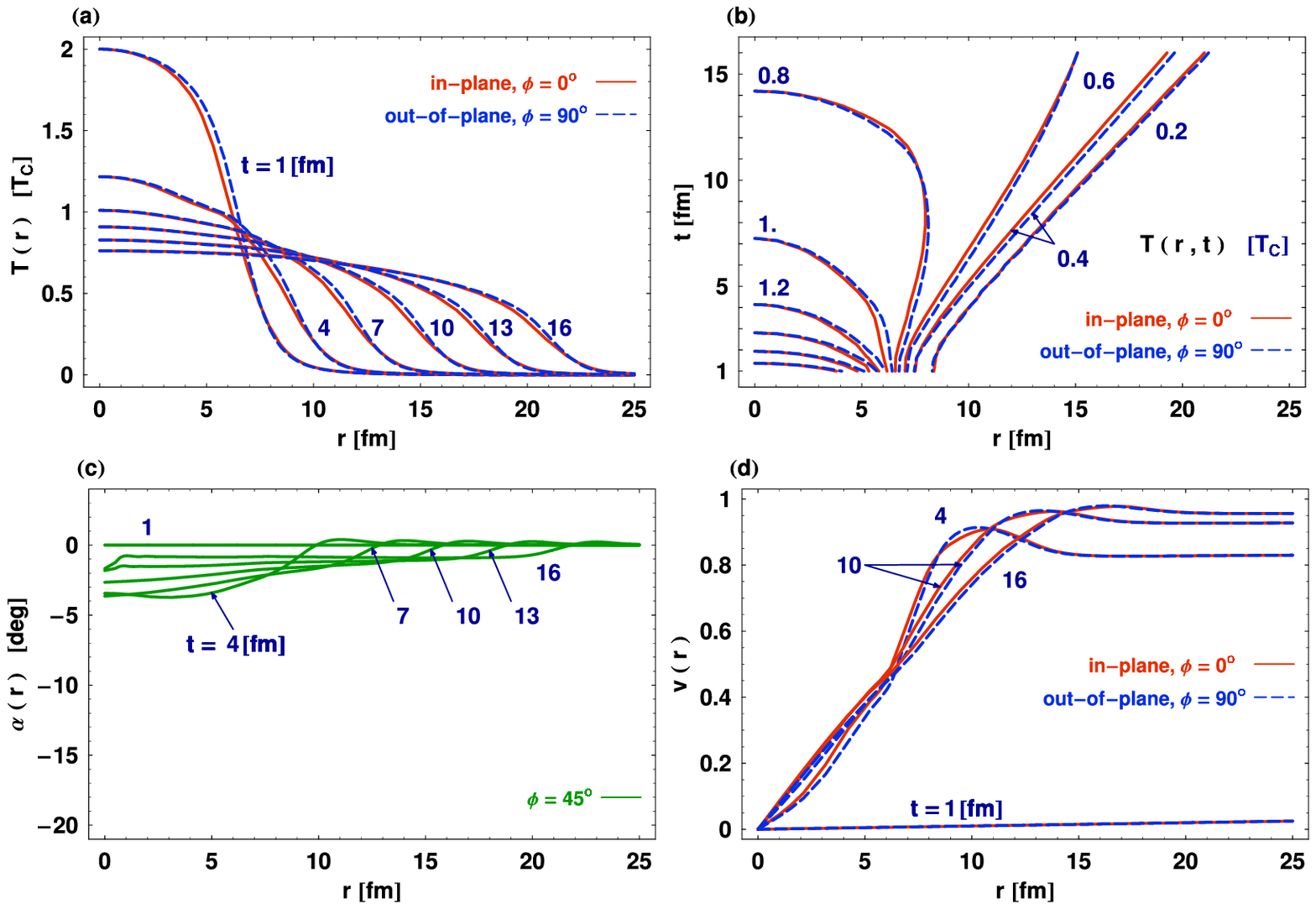}} \\
\subfigure{\includegraphics[angle=0,width=0.95\textwidth]{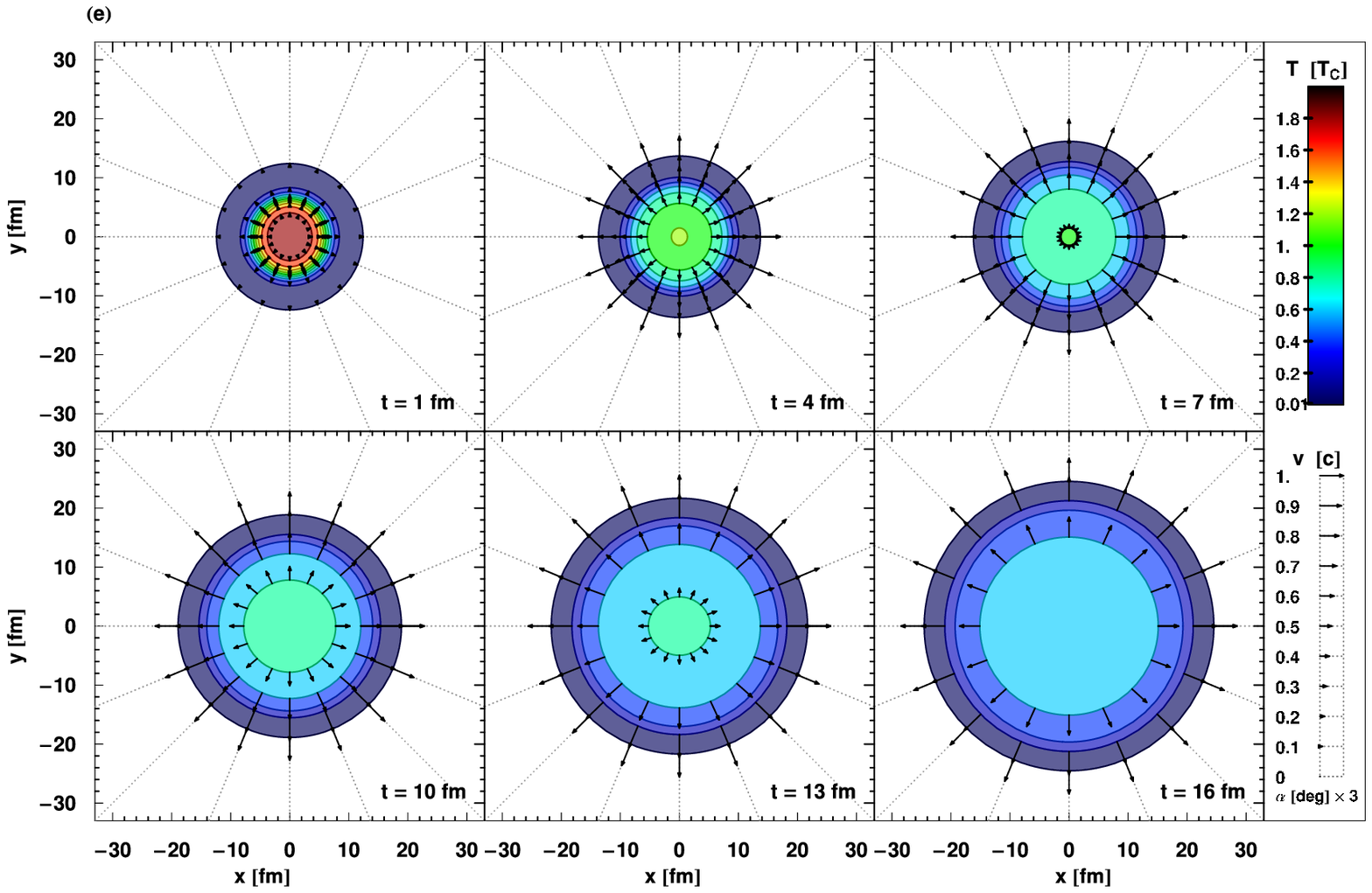}}
\end{center}
\caption{Time development of matter characterized by the initial conditions (\ref{initv}) -- (\ref{aT}) with $H_0 = 0.001 \hbox{fm}^{-1}$,  $T_0~=~T(t_0,0,0)~=~2~T_c$, and $b=3.9$ fm. }
\label{fig:res1}
\end{figure*}

\begin{figure*}[t!]
\begin{center}
\subfigure{\includegraphics[angle=0,width=0.95\textwidth]{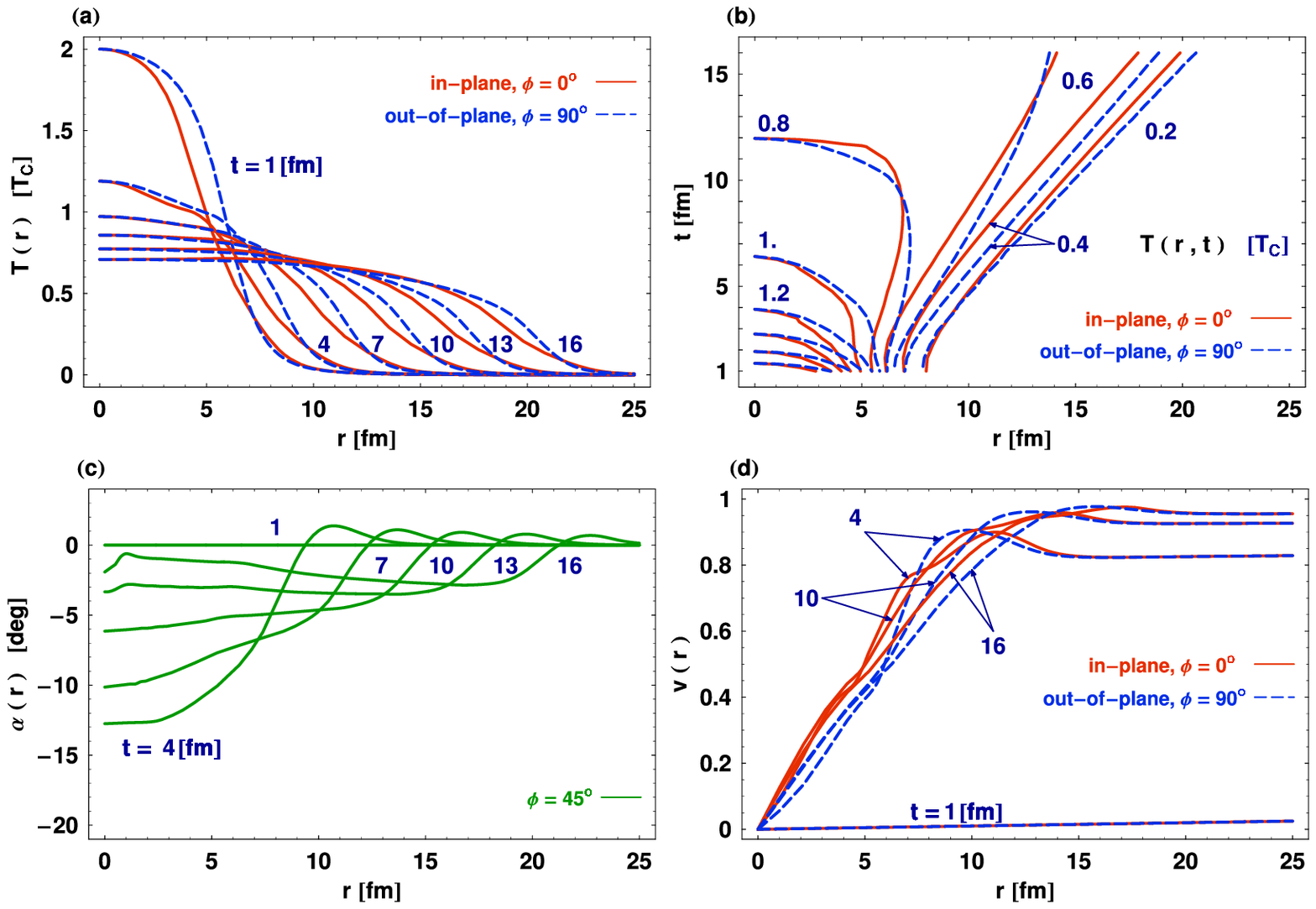}} \\
\subfigure{\includegraphics[angle=0,width=0.95\textwidth]{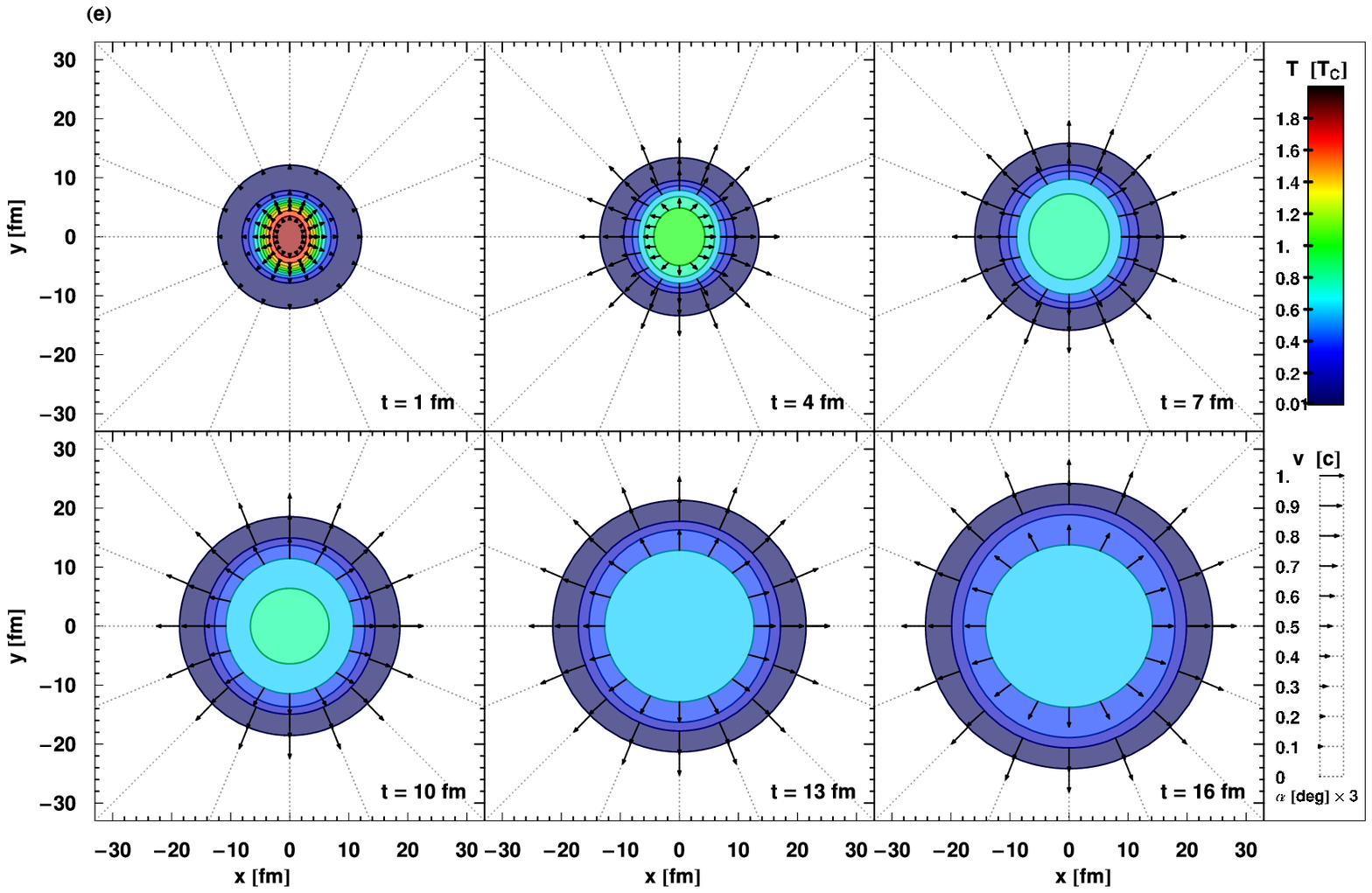}}
\end{center}
\caption{Same as Fig. \ref{fig:res1} but with $b=7.1$ fm. }
\label{fig:res2}
\end{figure*}

\begin{figure*}[t!]
\begin{center}
\subfigure{\includegraphics[angle=0,width=0.95\textwidth]{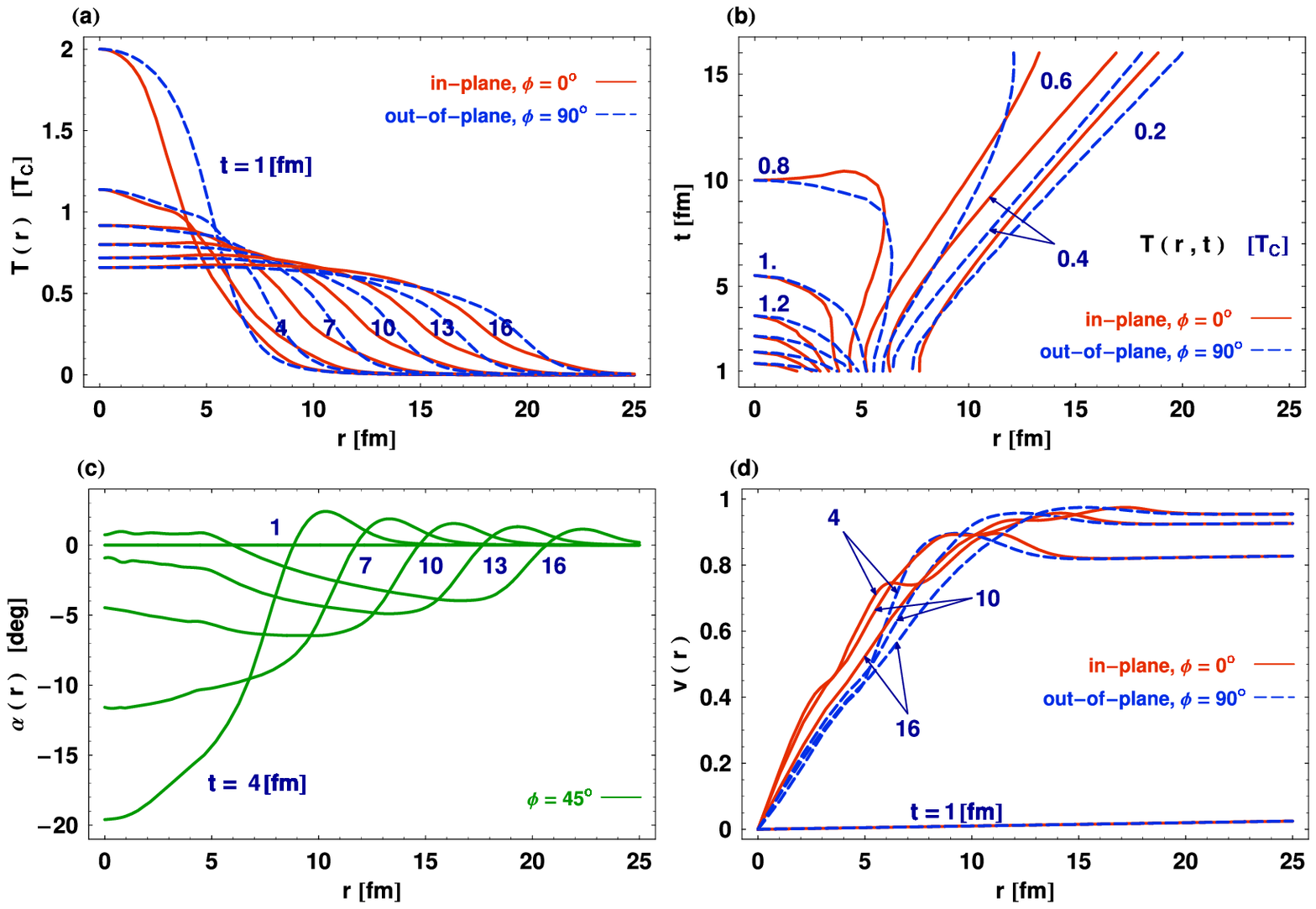}} \\
\subfigure{\includegraphics[angle=0,width=0.95\textwidth]{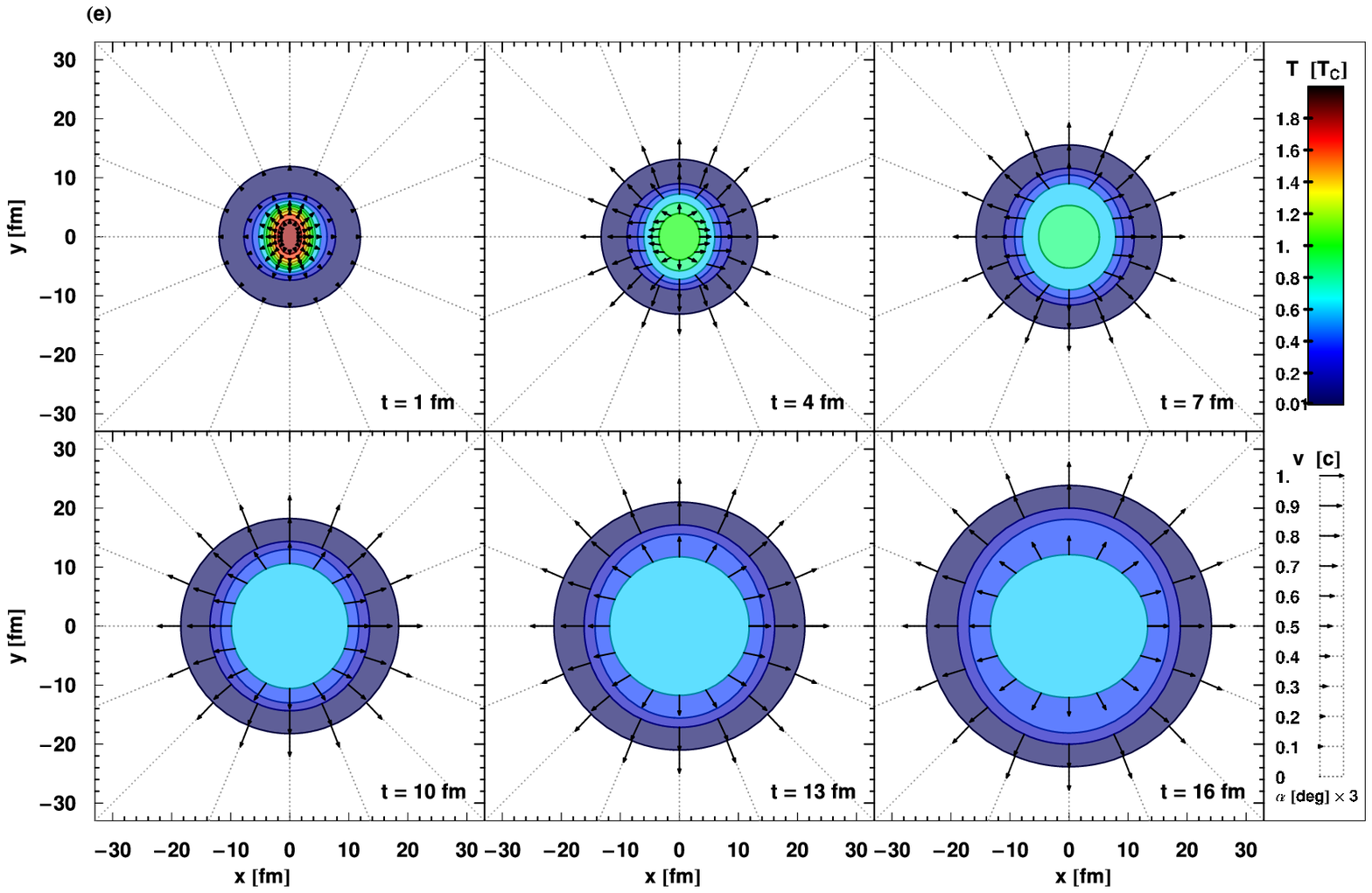}}
\end{center}
\caption{Same as Fig. \ref{fig:res1} but with $b=9.2$ fm. }
\label{fig:res3}
\end{figure*}

The examples of our results are shown in Figs.~\ref{fig:res1}~-~\ref{fig:res3}. The three sets of plots correspond to three different values of the impact parameter;  $b$ = 3.9, 7.1, and 9.2~fm. These values are typical for the three centrality classes: $c=0-20\%$, $c=20-40\%$, and $c=40-60\%$. The  relation between the (average) impact parameter and the centrality class is obtained from the formula,
\begin{equation}
b = \frac{1}{c_{\rm max}-c_{\rm min}} \int_{c_{\rm min}}^{c_{\rm max}}\!dc\, b(c) =
\frac{4 r_0}{3} \frac{c_{\rm max}^{3/2}-c_{\rm min}^{3/2}}{c_{\rm max}-c_{\rm min}},
\end{equation}
where the geometric relation between the centrality and the impact parameter is used, $c = b^2/(4 r_0^2)$ \cite{Broniowski:2001ei}. In all considered cases the initial central temperature equals twice the critical temperature, $T_0 = T(t_0,0,0) = 2 T_c$.

The parts a) of Figs. \ref{fig:res1} - \ref{fig:res3} show the temperature profiles for different values of time: $t = 1, 4, 7, 10, 13, 16$ fm. The solid lines represent the temperature profiles along the $x$-axis ($\phi=0$), while the dashed lines represent the profiles along the $y$-axis ($\phi=\pi/2$). One can observe that during the whole considered evolution time, 1 fm $ < t < $ 16 fm, the $y$-extension of the system remains larger than the $x$-extension.  However, the relative magnitude of this effect decreases with time, indicating that the cylindrical symmetry is gradually restored as the time increases. In these parts of the plots one can also notice the effect of the phase transition; initially the system cools down rather rapidly, later the cooling down is delayed and the main visible effect is the increase of the volume of the system. We note that this behavior is related to the sudden decrease of the sound velocity in the region $T \approx T_c$.

The parts b) of Figs. \ref{fig:res1} - \ref{fig:res3} show the isotherms in the $t-r$ space, again for $\phi=0$ (solid lines) and $\phi=\pi/2$ (dashed lines). The pairs of isotherms indicate different values of the temperature. They start at $T = 1.8 \,T_c$ and go down to $T = 0.2 \,T_c$, with a step of $0.2 \,T_c$. It is interesting to observe (especially for $b$ = 7.1~fm and 9.2~fm) that the solid and dashed lines cross each other. This effect means that the central (relatively hotter) part of the system acquires a pumpkin-like shape during the evolution of the system. Such pumpkin-like regions, however, shrink and disappear during further expansion. 

In the parts c) of Figs.~~\ref{fig:res1}~-~\ref{fig:res3} the profiles of the function $\alpha(t,r,\phi)$ are shown for $t = 1, 4, 7, 10, 13, 16$ fm and $\phi = \pi/4$. For $\phi=0$ and $\phi=\pi/2$ the function $\alpha(t,r,\phi)$ vanishes due to the symmetry reasons. In the first and third quadrant the values of $\alpha$ are predominantly negative, while in the second and fourth quadrant the values of $\alpha$ are positive. This behavior characterizes the direction of the flow which has the tendency to change the initial almond shape into a cylindrically symmetric shape. We have checked that positive values of $\alpha$ at $\phi = \pi/4$, seen at the edge of the system, are caused by the overlapping tails of the Woods-Saxon distributions.

In the parts d) the velocity profiles are shown, again for $t$ = 1, 4, 10, 16 fm. Similarly to the previously discussed figures, the solid lines are the profiles for $\phi=0$ ($x$-direction), whereas the dashed lines are the profiles for $\phi=\pi/2$ ($y$-direction). One can observe that the magnitude in the $x$-direction is larger than the magnitude in the $y$-direction, which is an expected hydrodynamic behavior caused by larger pressure gradients in the $x$-direction. Exactly this effect is responsible for the observed azimuthal asymmetry of the transverse-momentum spectra, quantified by the values of the $v_2$ coefficient.

Finally, in the parts e) we show the contour lines of the temperature, again for different values of time. These plots visualize the time development of the system. The arrows describe the magnitude and the direction of the flow (for better recognition the angle $\alpha$ is magnified by a factor of 3).

\section{Summary}

In this paper we introduced a new and concise treatment of the boost-invariant and cylindrically non-symmetric relativistic hydrodynamic equations. This approach may be used to describe the evolution of matter produced in the central region of ultra-relativistic heavy-ion collisions, such as investigated in the present RHIC or future LHC experiments. The presented formalism is a direct generalization of the approach introduced by Baym et al. in Ref. \cite{Baym:1983sr}. In the studied case, the symmetry of the problem allows us to rewrite the hydrodynamic equations as only two coupled differential equations, (\ref{alphaeq}) and (\ref{aeq}), which automatically lead to the fulfillment of the requested boundary conditions for the velocity and the temperature at the center of the system. The effects of the phase transition are included in this scheme by the use of the temperature dependent sound velocity. As long as the sound velocity satisfies the condition against the shock formation, the resulting hydrodynamic equations may be solved numerically with the help of the standard methods. We note that besides the sound velocity, no other thermodynamic variables are necessary for solving the hydrodynamic equations in this case. The presented results of the hydrodynamic calculations, supplemented with the appropriate freeze-out model,  may be used to calculate different physical observables. In particular, the coefficient of the elliptic-flow, $v_2$, may be extracted. To achieve this task, in the closest future we intend to combine our hydrodynamic approach with the statistical Monte-Carlo model {\tt THERMINATOR} \cite{Kisiel:2005hn}.

\end{document}